\documentclass[runningheads]{llncs}

\usepackage[T1]{fontenc}
\usepackage{graphicx}
\usepackage{booktabs}
\usepackage{textcomp}
\usepackage{verbatim,enumitem}
\usepackage{algorithm}
\usepackage{colortbl}
\usepackage{multirow}
\usepackage{subcaption}
\usepackage[utf8]{inputenc}
\usepackage{lipsum}
\usepackage{framed}
\usepackage{listings}
\usepackage{url}
\usepackage{enumitem}
\usepackage{float}
\usepackage{hyperref}          %
\usepackage{xurl}              %
\hypersetup{breaklinks=true}
\usepackage[capitalise,noabbrev]{cleveref} 
\crefname{appsec}{Appendix}{Appendices}

\lstdefinestyle{python}{
    language=Python,
    basicstyle=\ttfamily\footnotesize,
    keywordstyle=\color{blue},
    stringstyle=\color{green!40!black},
    commentstyle=\color{gray},
    showstringspaces=false,
    breaklines=true,
    frame=single,
    captionpos=b,
    numbers=left,
    numberstyle=\tiny\color{gray},
    stepnumber=1,
    numbersep=5pt,
}

\begin{document}

\title{Prompt and Circumstances: Evaluating the Efficacy of Human Prompt Inference in AI-Generated Art}
\titlerunning{Prompt and Circumstances}

\author{
Khoi Trinh\inst{1} \and
Scott Seidenberger\inst{1} \and
Joseph Spracklen\inst{2} \and
Raveen Wijewickrama\inst{2} \and
Bimal Viswanath\inst{3} \and
Murtuza Jadliwala\inst{2} \and
Anindya Maiti\inst{1}
}

\authorrunning{Trinh et al.}

\institute{
University of Oklahoma, Norman, OK, USA\\
\email{khoitrinh@ou.edu, seidenberger@ou.edu, am@ou.edu}
\and
University of Texas at San Antonio, San Antonio, TX, USA\\
\email{joseph.spracklen@my.utsa.edu, raveen.wijewickrama@utsa.edu, murtuza.jadliwala@utsa.edu}
\and
Virginia Polytechnic Institute and State University, Blacksburg, VA, USA\\
\email{vbimal@vt.edu}
}

\maketitle
\begin{abstract}

The emerging field of AI-generated art has witnessed the rise of prompt marketplaces, where creators can purchase, sell, or share prompts to generate unique artworks. These marketplaces often assert ownership over prompts, claiming them as intellectual property. This paper investigates whether concealed prompts sold on prompt marketplaces can be considered as bona fide 
intellectual property, given that humans and AI tools may be able to sufficiently 
infer the prompts based on publicly advertised sample images accompanying each prompt on sale. 
Specifically, our study 
aims to assess (i) how accurately humans can infer the original prompt solely by examining an AI-generated image, with the goal of generating images similar to the original image, and (ii) the possibility of improving upon individual human and AI 
prompt inferences by crafting human-AI combined prompts with the help of a large language model (LLM).
Although previous research has explored AI-driven prompt inference and protection strategies, our work is the first to incorporate a human subject study and examine collaborative human-AI prompt inference in depth.
Our findings indicate that while prompts inferred by humans and prompts inferred through a combined human-AI effort
can generate images with a moderate level of similarity, they are not as successful as using the original prompt. Moreover, combining human- and AI-inferred prompts using our suggested merging techniques did not improve performance over purely human-inferred prompts.

\end{abstract}

\section{Introduction}
\label{sec:introduction}

Artificial Intelligence (AI) has transformed creative expression by enabling anyone to generate visually rich and conceptually coherent artworks from text prompts. Central to this process are text-to-image (\texttt{txt2img}) models such as Midjourney~\cite{midjourney}, DALL-E~\cite{ramesh2021zero}, Stable Diffusion~\cite{rombach2022high}, and GLIDE~\cite{nichol2022glide}. These systems combine a language encoder (e.g., CLIP~\cite{clip}) with a diffusion-based image generator trained on large text-image datasets (e.g., LAION-5B~\cite{schuhmann2022laion}), translating textual descriptions into high-quality images.

Prompts are the primary means of steering such models and are often crafted through iterative trial and error~\cite{wang2023review}. Their effectiveness has given rise to professional ``prompt engineering'' and online marketplaces where creators buy and sell carefully designed prompts that yield distinctive artistic styles. Because well-crafted prompts are difficult to reproduce, many platforms treat them as proprietary information.

Recent copyright guidance clarifies that fully AI-generated works lack protection, but human–AI collaborations may qualify if they include meaningful human authorship~\cite{us2023copyright,usco2025report,thaler2025}. Consequently, prompt marketplaces assert intellectual property rights over prompts~\cite{promptbasetos,promptrrtos}, raising new questions about ownership and vulnerability to \emph{prompt inference}—the act of reconstructing or approximating an original prompt from its generated image. Prior work has discussed automated prompt stealing attacks and countermeasures~\cite{shen2023prompt,struppek2022rickrolling,zhai2023text}, yet little is known about how accurately humans or human–AI teams can perform such inference.

This work investigates three main research questions: \emph{(i) how accurately can humans infer the text prompt behind an AI-generated image, (ii) to what extent do factors such as prompt length, complexity, or image generation model influence prompt inference results, and (iii) can large language models assist humans in producing prompts that yield visually similar results?} To address them, we conduct a human-subject study assessing participants' ability to infer prompts from AI-generated images and introduce a novel distribution-level evaluation method based on the Kolmogorov–Smirnov (KS) test to determine when two prompts produce statistically indistinguishable image distributions. This offers a rigorous alternative to single-image similarity comparisons and better accounts for the stochasticity of generative models.

Our results show that while humans can partially infer the subjects of prompts, accurately capturing stylistic modifiers remains difficult. AI-augmented human inferences do not outperform purely human attempts, suggesting that prompt-based intellectual property remains relatively resilient to reverse engineering under current techniques.

\section{Background and Related Work}
\label{sec:related}

\subsection{Prompt Marketplaces and Prompt Inference in AI Art}
\label{sec:back-market}

\begin{figure}[htbp]
\centering
\includegraphics[width=0.8\linewidth]{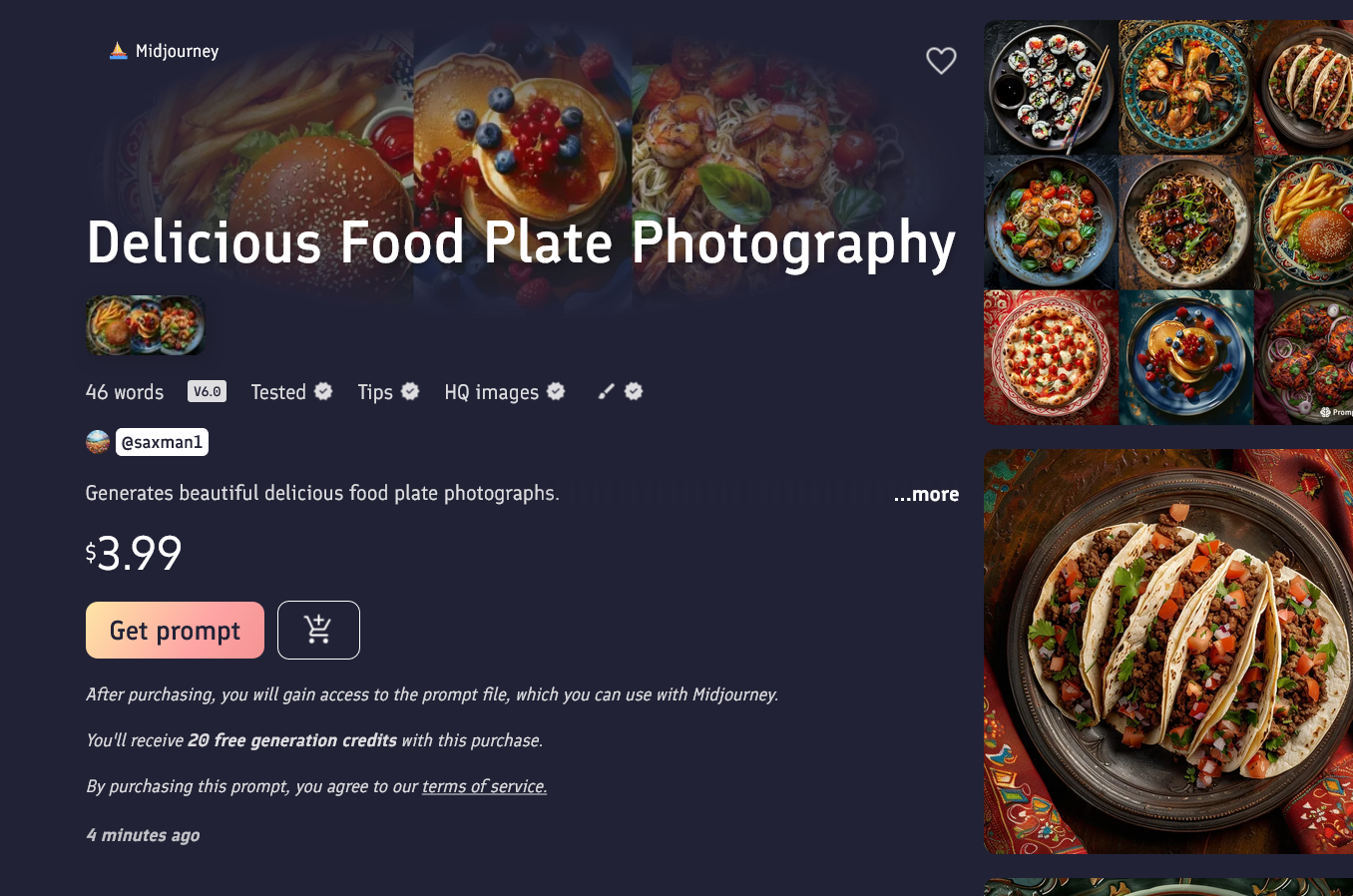}
\caption{Example PromptBase listing selling a prompt associated with a specific artistic style.}
\label{fig:promptbase-example}
\end{figure}

The popularity of \texttt{txt2img} models has spurred the rise of prompt marketplaces such as PromptBase, PromptHero, and CivitAI, where users buy, sell, and share text prompts for generating high-quality AI art (\cref{fig:promptbase-example}). These platforms operate on commission-based models and treat prompts as creative intellectual property~\cite{promptbasetos,promptrrtos}, recognizing their role as both artistic tools and commercial assets. Effective prompts substantially improve image quality~\cite{decrypt.coLucrative,makeuseofAIprompts}, leading to a small but growing economy of ``prompt engineering'' professionals and collectors.

\subsection{Prompt Inference, Subjects, and Modifiers}
\label{sec:back-subjects-modifiers}

Prior work has examined both the vulnerability of prompts to inference attacks and methods to protect them. Several studies have shown that machine learning models can reverse-engineer or approximate hidden prompts~\cite{clipinterrogator,shen2023prompt,wu2022membership,li2022blip}, while others propose defensive techniques such as backdoor injection, data poisoning, and anomaly detection~\cite{struppek2022rickrolling,zhai2023text}. Yet these efforts focus mainly on automated systems rather than on human capabilities. If humans can accurately infer prompts through visual reasoning, the practical value of such defenses may be limited.

Recent work on human-AI co-creation~\cite{lyu2022communication,brade2023promptify} suggests that collaborative prompting may enhance creativity, but its potential for reconstructing prompts has not been studied. Understanding this human-AI interplay is critical for assessing whether collaborative inference threatens the originality or ownership of text prompts.

Prompts generally consist of a \emph{subject}, defining the image’s central theme (e.g., ``cat,'' ``forest,'' ``robot''), and one or more \emph{modifiers}, describing stylistic or contextual details such as lighting, texture, emotion, or medium. These linguistic components jointly steer the generative model toward specific visual outcomes. As shown in \cref{fig:sub-mod-example}, even small modifier changes can substantially alter the resulting imagery, underscoring the complexity of prompt interpretation and inference.

\section{Research Questions \& Hypotheses}
\label{sec:goals}

\begin{figure}[htbp]
\centering
\includegraphics[width=0.8\linewidth]{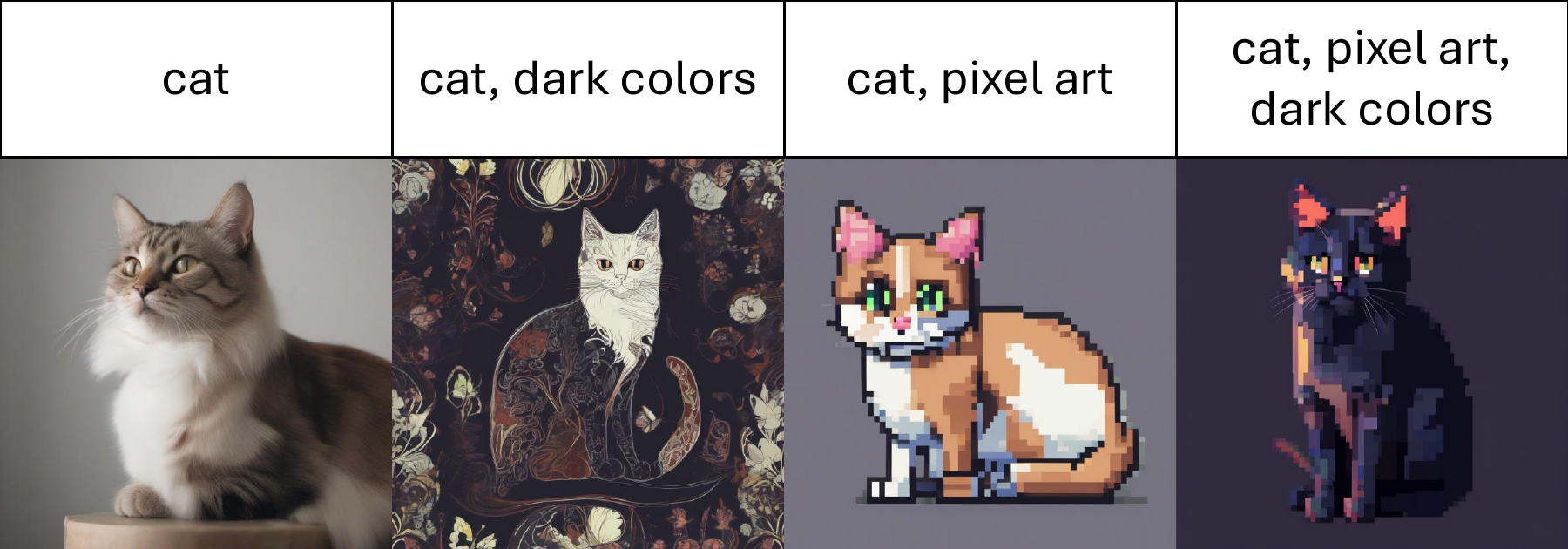}
\caption{Image generations using SDXL with prompts containing the same subject (cat) and two modifiers (pixel art, dark colors).}
\label{fig:sub-mod-example}
\end{figure}

Inferring prompts behind AI-generated images is challenging for both humans and AI models~\cite{clipinterrogator,img2prompt,shen2023prompt,lu2023seeing,liu2022design} due to the complex interplay between visual content and linguistic nuance. \cref{fig:sub-mod-example} illustrates how subtle modifier changes for the same subject can yield markedly different images. Humans rely on subjective interpretation when describing such content, often diverging in how they reconstruct subjects and stylistic cues. Understanding this human reasoning—alone or with AI support—remains largely unexplored and motivates our study.

A \textbf{key methodological contribution} of this work is a principled way to determine when two prompts generate statistically indistinguishable images. We treat each prompt as a \emph{distribution} of outputs rather than a single instance, drawing multiple samples across seeds and comparing the resulting similarity-score distributions. Using the two-sample Kolmogorov–Smirnov (KS) test, we test whether these distributions differ significantly. This non-parametric, model-agnostic framing captures stochastic variability in \texttt{txt2img} models and extends prior single-image heuristics. It aligns with two-sample testing approaches such as kernel-MMD and classifier-based tests~\cite{gretton2012kernel,lopez2016revisiting}. To our knowledge, \textbf{formalizing prompt inference as a distributional equivalence problem and evaluating it with the KS test is novel in this domain}. Details appear in \cref{sec:success-scores-expsetup} and \cref{fig:hit-miss-examples}.

\paragraph{Research Questions.}
\begin{itemize}
    \item \textbf{RQ1}: To what extent can humans accurately infer the original text prompts used to generate AI art, based solely on the visual content of the images?
    \item \textbf{RQ2}: How do specific factors such as subject complexity, modifier ambiguity, or image generation model influence prompt inference performance?
    \item \textbf{RQ3}: Can large language models (LLMs) improve human prompt inference through collaborative, combined prompt construction?
\end{itemize}

\paragraph{Hypotheses.}
\begin{itemize}
    \item \textbf{H1}: Humans can infer prompts with moderate success, especially recognizing subjects, but struggle with nuanced modifiers~\cite{lu2023seeing}.
    \item \textbf{H2}: Inference success correlates with prompt characteristics (length, adjective and clause counts) and the underlying generation model, consistent with prior findings that richer descriptions aid interpretation~\cite{moessner2024human}.
    \item \textbf{H3}: Combining human-inferred prompts with AI-generated suggestions (e.g., CLIP Interrogator + GPT-4) will yield higher similarity scores than human inference alone~\cite{brade2023promptify,mahdavi2024ai}.
\end{itemize}

These questions and hypotheses inform our study design and evaluation (\cref{sec:survey,sec:success-scores-expsetup}), enabling us to examine not only the mechanics of human and AI prompt inference but also its implications for prompt ownership and protection in commercial marketplaces.

\section{Study Design and Participants} 
\label{sec:survey}

\subsection{Prompt and Image Datasets}
\label{sec:dataset}

We generated sample images from two prompt types: \textit{Controlled} and \textit{Uncontrolled}. The two sets differ in the structure and constraint of the prompts, not in the availability of ground-truth text. In both cases, the original prompts are known to the authors and are used to generate the target images' ISM distributions for evaluation.

Controlled prompts provided a baseline for evaluating recognition of common subjects and modifiers across models and participant groups, while uncontrolled prompts introduced greater diversity and served as open-ended inference tasks.

\subsubsection{Controlled Dataset.}
\label{sec:controlled-prompts}

The controlled set was built from prompts scraped from Lexica\footnote{\url{https://lexica.art}}, a public prompt-sharing platform. Using a \textit{Selenium}-based script and \textit{spaCy} parsing, we analyzed over 100,000 prompts to identify frequent subjects: \textit{man}, \textit{woman}, \textit{astronaut}, \textit{cat}, and \textit{robot}. We then paired these with 121 stylistic modifiers drawn from Midjourney’s keyword repository~\cite{midjourneymodifiers}, spanning lighting, mood, medium, color palette, and perspective categories. Random combinations (1–5 modifiers per subject) yielded 100 controlled prompts.

We refer to this dataset as \emph{controlled} because both the subject and the modifier components of each prompt are explicitly known and constrained by the experimental design. This structure enables direct comparison between the original prompt and participant-inferred prompts, and facilitates analysis of which prompt components (e.g., subjects versus modifiers) are more readily recovered.

\subsubsection{Uncontrolled Dataset.}
\label{sec:uncontrolled-prompts}

The uncontrolled set consisted of 100 complete prompts randomly sampled from PromptHero, covering a wide stylistic range. These prompts were used to generate images for Part~II of the survey, 25 per model across four \texttt{txt2img} systems.

\subsection{Analyzed \texttt{txt2img} Models}
\label{sec:txt2img}

We focused on four widely used models representing both base and fine-tuned architectures:

\begin{itemize}[leftmargin=*]
\item \textbf{MidJourney v5.0}~\cite{midjourney}: closed-source diffusion model producing highly stylized 1024$\times$1024 images via Discord interface.
\item \textbf{Stable Diffusion XL (SDXL)}~\cite{podell2023sdxl,rombach2022high}: open-source transformer-based model capable of 1024$\times$1024 generation.
\item \textbf{DreamShaper XL}: fine-tuned on SDXL for photo-realistic fantasy imagery, maintaining native 1024$\times$1024 resolution.
\item \textbf{Realistic Vision v5}~\cite{rv5}: fine-tuned on Stable Diffusion 1.5 for realistic photography; outputs were upscaled from 512$\times$512 to 1024$\times$1024 using \textit{Highres.fix} and ScuNET-PSNR for fair comparison.
\end{itemize}

All target images and participant-submitted prompt generations were produced with the following settings shown in \cref{appendix-gen-params}.

\begin{table}[htbp]
  \centering
  \resizebox{\linewidth}{!}{%
  \begin{tabular}{@{}ll@{}}
    \toprule
    \textbf{Parameter} & \textbf{Value} \\
    \midrule
    Resolution (Stable Diffusion XL and DreamShaper XL) & 1024 $\times$ 1024 \\
    Resolution (Realistic Vision 5) & 512 $\times$ 512 $\rightarrow$ 1024 $\times$ 1024 (upscaled) \\
    CFG Scale & 5 \\
    Sampling Steps & 40 \\
    Sampling Method & Euler \\
    \midrule
    \multicolumn{2}{@{}l}{\textbf{Upscaling Settings for RV5}} \\
    \quad Upscaler & ScuNET PSNR \\
    \quad Denoising Strength & 0.4 \\
    \quad Restore Faces & On \\
    \quad Hires. Fix & On \\
    \midrule
    Midjourney & Parameters not user-controllable; \\
           & generations used default model settings \\
    \bottomrule
  \end{tabular}
  }
  \caption{Generation parameters used across models.}
  \label{appendix-gen-params}
\end{table}

At the study’s start (late 2023), these models were the most active on major prompt marketplaces. DALL-E was excluded due to frequent version changes that affected replicability.

\subsection{Pre-Survey: Demographics and Familiarity}

\begin{figure}[htbp]
\centering
\begin{subfigure}[b]{0.4\linewidth}
    \centering
    \includegraphics[width=\textwidth]{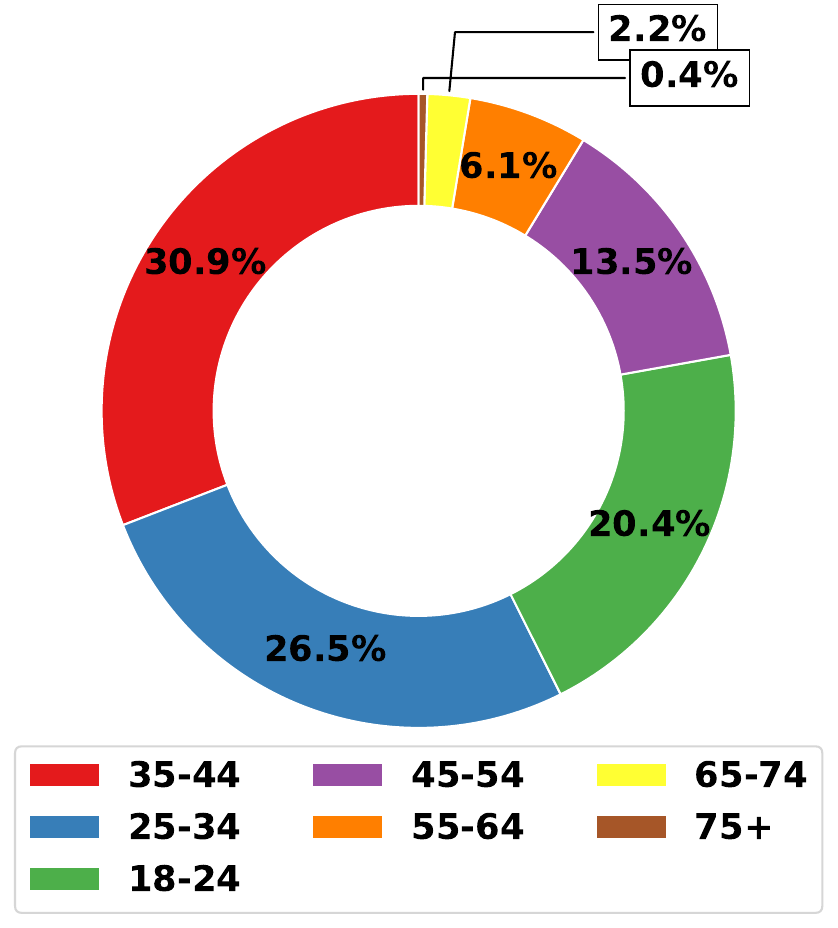}
    \caption{Age distribution.}
    \label{fig:age-distribution}
\end{subfigure}
\quad
\begin{subfigure}[b]{0.4\linewidth}
    \centering
    \includegraphics[width=\textwidth]{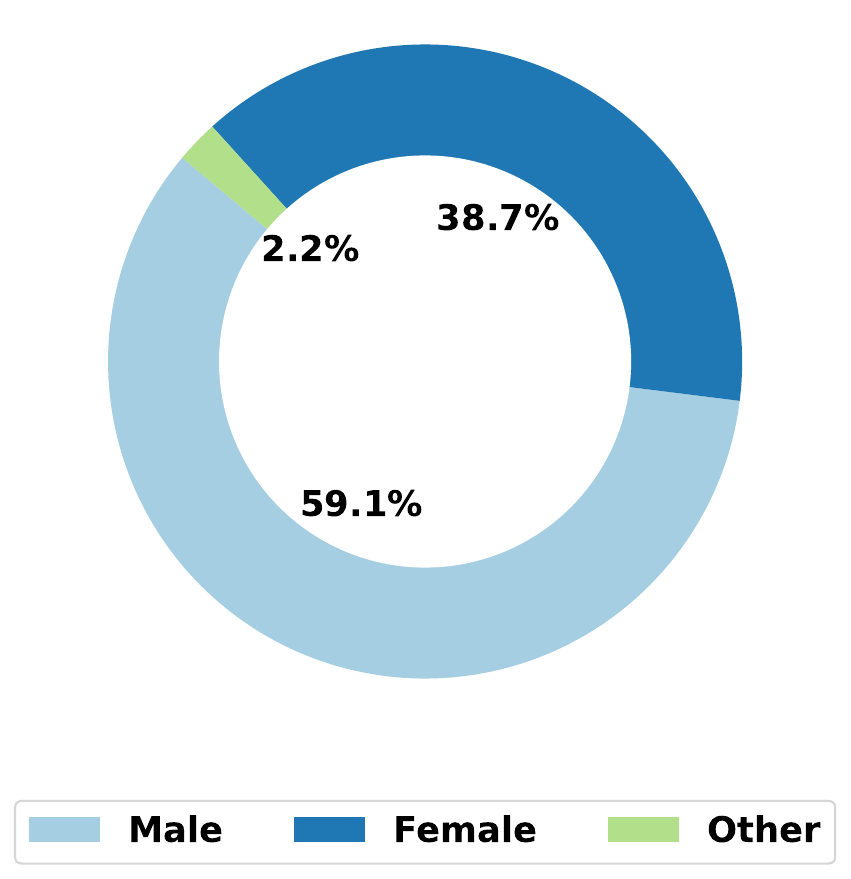}
    \caption{Gender distribution.}
    \label{fig:gender-distribution}
\end{subfigure}
\caption{Demographic distribution of survey participants.}
\label{fig:demographics}
\end{figure}

A total of 230 participants (on-campus and online via Amazon Mechanical Turk) completed the survey between December 2023 and March 2024. Most were aged 18–44, with a slight male majority (59\%), and 16\% reporting an arts background. \cref{fig:demographics} visualizes this breakdown. 

\begin{figure}[htbp]
\centering
\includegraphics[width=0.8\linewidth]{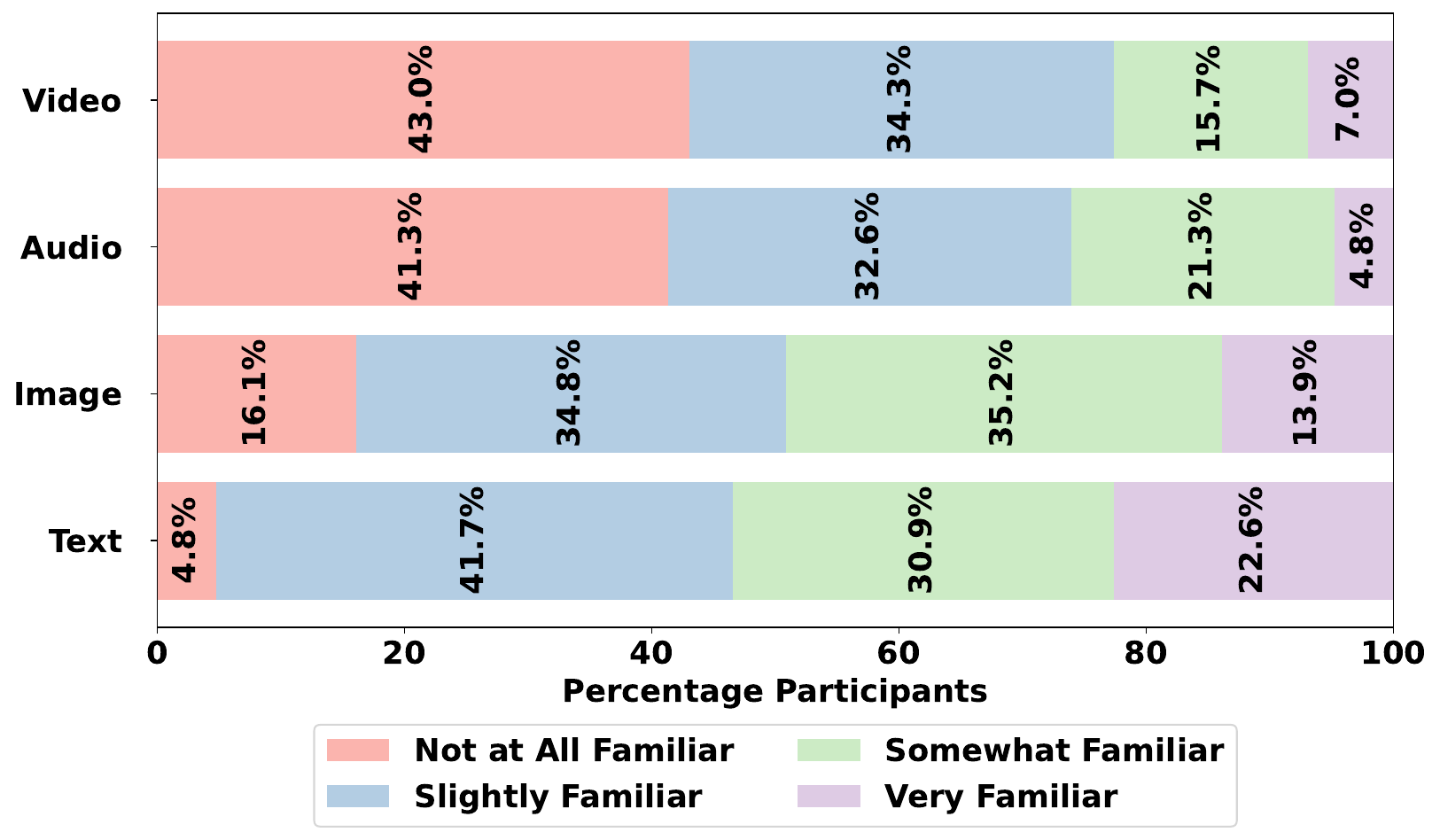}
\caption{Participants' familiarity levels with different generative AI tools. } 
\label{fig:ai-tool-familiarity}
\end{figure}

Participants’ familiarity with generative AI tools was moderate overall: roughly one-third were ``\emph{Slightly Familiar}'' with image generation tools, with similar familiarity levels across text, audio, and video applications (\cref{fig:ai-tool-familiarity}). This mix provided a balanced range of prior experience for evaluating prompt inference performance.

\subsection{Survey Design and Flow}
\label{sec:survey-design}

\begin{figure}[htbp]
  \centering
  \includegraphics[width=0.8\linewidth]{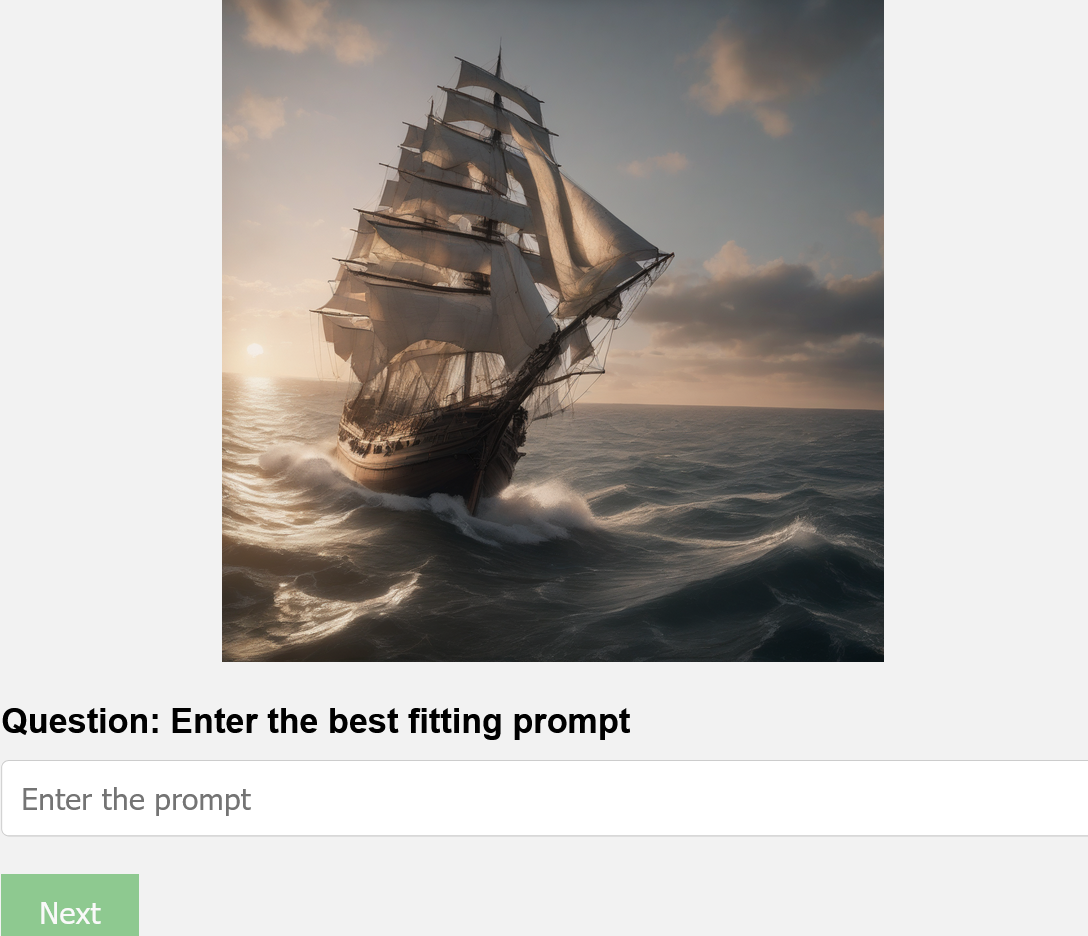}
  \vspace{-0.1in}
  \caption{An example of a survey question where the participant must type in a prompt for the given target image.}
  \vspace{-0.1in}
  \label{fig:p3}
\end{figure}

For our main user study to assess the ability of humans to infer prompts from AI-generated images, we design and implement a custom tool for our study interface.
This tool dynamically loads and displays images together with the instructions, and can capture participant responses to these target images. \cref{fig:p3} shows an example of a question that participants saw.

\textbf{Part~I (Uncontrolled Dataset).} Participants viewed five AI-generated images and wrote full-sentence prompts describing each, specifying both subject and modifiers. This task measured open-ended inference and linguistic creativity.

\textbf{Part~II (Controlled Dataset).} Participants viewed another five images from the controlled set and generated descriptive prompts. Because the true prompts were known, these responses enabled direct comparison of inferred versus original text. An example interface is shown in \cref{fig:p3}.

Responses were filtered post-survey: entries were excluded if they lacked a subject or modifier, contained random or blank text, or were non-English. After filtering, we retained 1,141 valid responses for controlled and 1,145 for uncontrolled sets, forming the basis for subsequent evaluation.

\section{Experimental Setup}
\label{sec:exp-setup}

\subsection{Image Similarity Metrics}
\label{sec:metrics}

We evaluate prompt inference accuracy by comparing the images generated from participants’ inferred prompts to those from the original prompts using several image similarity metrics (ISMs). These metrics capture both perceptual and semantic relationships, providing a comprehensive measure of how closely inferred prompts reproduce the intended visuals.

\subsubsection{Image Hash}
We use the \texttt{imagehash} library\footnote{\url{https://pypi.org/project/ImageHash/}} to compute perceptual hashes and compare them via Hamming distance. A smaller distance indicates higher similarity between the image generated from the inferred prompt and the original reference image. This metric yields a discrete similarity score ranging from 0 (most similar) to 64 (least similar).

\subsubsection{Learned Perceptual Image Patch Similarity (LPIPS)}
LPIPS~\cite{zhang2018perceptual} quantifies perceptual similarity in deep feature space, reflecting human visual judgments more accurately than pixel-level metrics. Lower LPIPS values indicate greater resemblance between compared images in terms of structure, texture, and color composition. This metric yields a continuous similarity score from 0 (most similar) to 1 (least similar).

\subsubsection{Image Embedding Similarity (CLIP Score)} 
The CLIP score~\cite{wang2023exploring} measures cosine similarity between text and image embeddings produced by OpenAI’s ViT-L/14 and ViT-B/32 CLIP models\footnote{\url{https://huggingface.co/openai/clip-vit-large-patch14}}. Higher scores imply stronger text–image alignment. We use SentenceTransformers~\cite{reimers2019sentence} wrappers for consistent embedding generation and joint computation of semantic and visual similarity. This metric yields a continuous similarity score from 0 (least similar) to 1 (most similar).

\subsection{Quantifying Successful Inferences}
\label{sec:success-scores-expsetup}

To account for the stochasticity of \texttt{txt2img} models, we evaluate prompt inference at the distribution level rather than per-image. For each prompt, we generate 200 reference images (original prompt) and 50 inferred images (participant prompt) across random seeds, treating each set as a sample from an underlying distribution.

We then compare the similarity-score distributions of inferred versus reference generations using the \textbf{two-sample Kolmogorov–Smirnov (KS) test}~\cite{massey1951kolmogorov,smirnov1948table}. The KS test, being non-parametric and sensitive to both distribution shape and location, robustly captures differences across generative variability. A prompt inference is considered a \textbf{hit} if no statistically detectable difference is observed between the two distributions at the 95\% confidence level ($p>0.05$)~\cite{fisher1970statistical}. Importantly, we do not treat ``hit" cases as evidence that two prompts are interchangeable; rather, we interpret these cases as the prompt differences not readily detectable under our evaluation procedure. The \textbf{hit rate} is then defined as the proportion of such successful inferences.

\begin{figure}[htbp]
    \centering
    \resizebox{\linewidth}{!}{%
    \begin{subfigure}[b]{0.24\textwidth}\includegraphics[width=\linewidth]{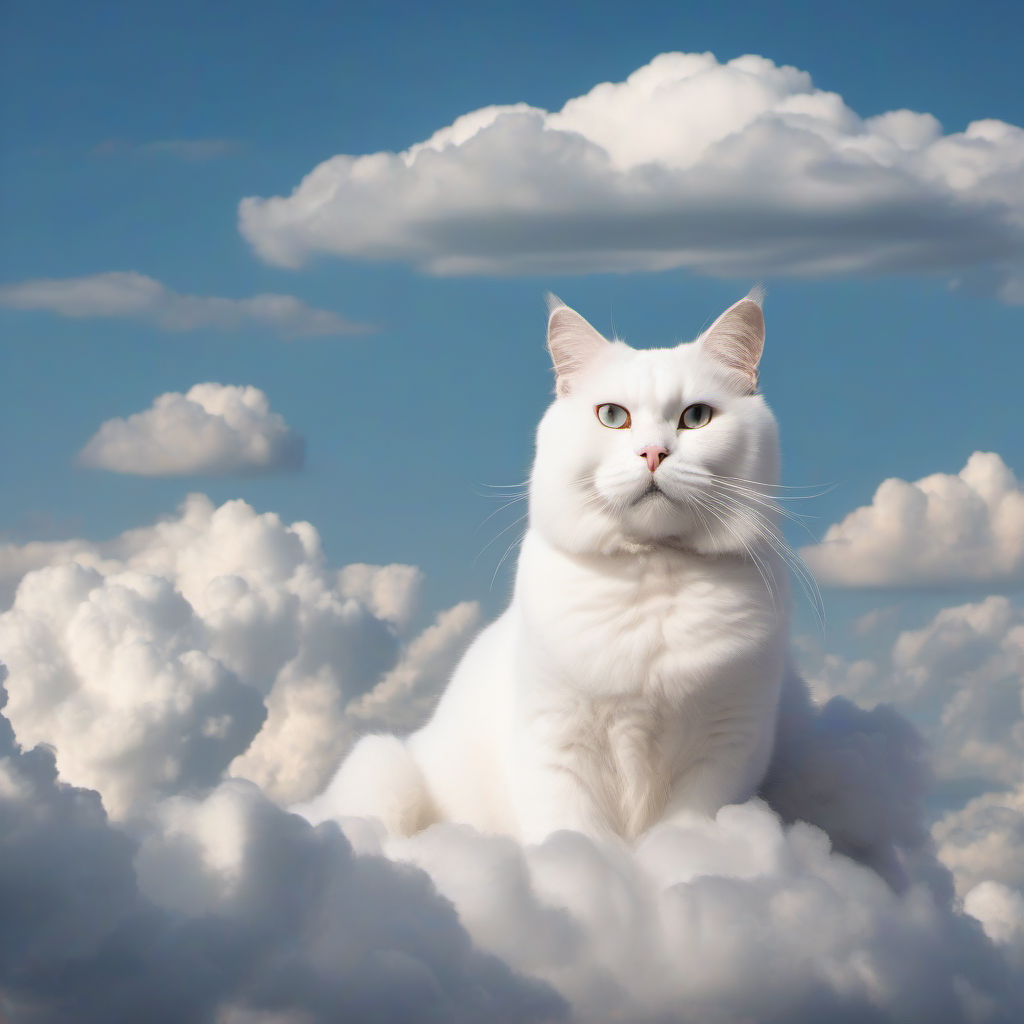}\caption{}\end{subfigure}
    \hfill
    \begin{subfigure}[b]{0.24\textwidth}\includegraphics[width=\linewidth]{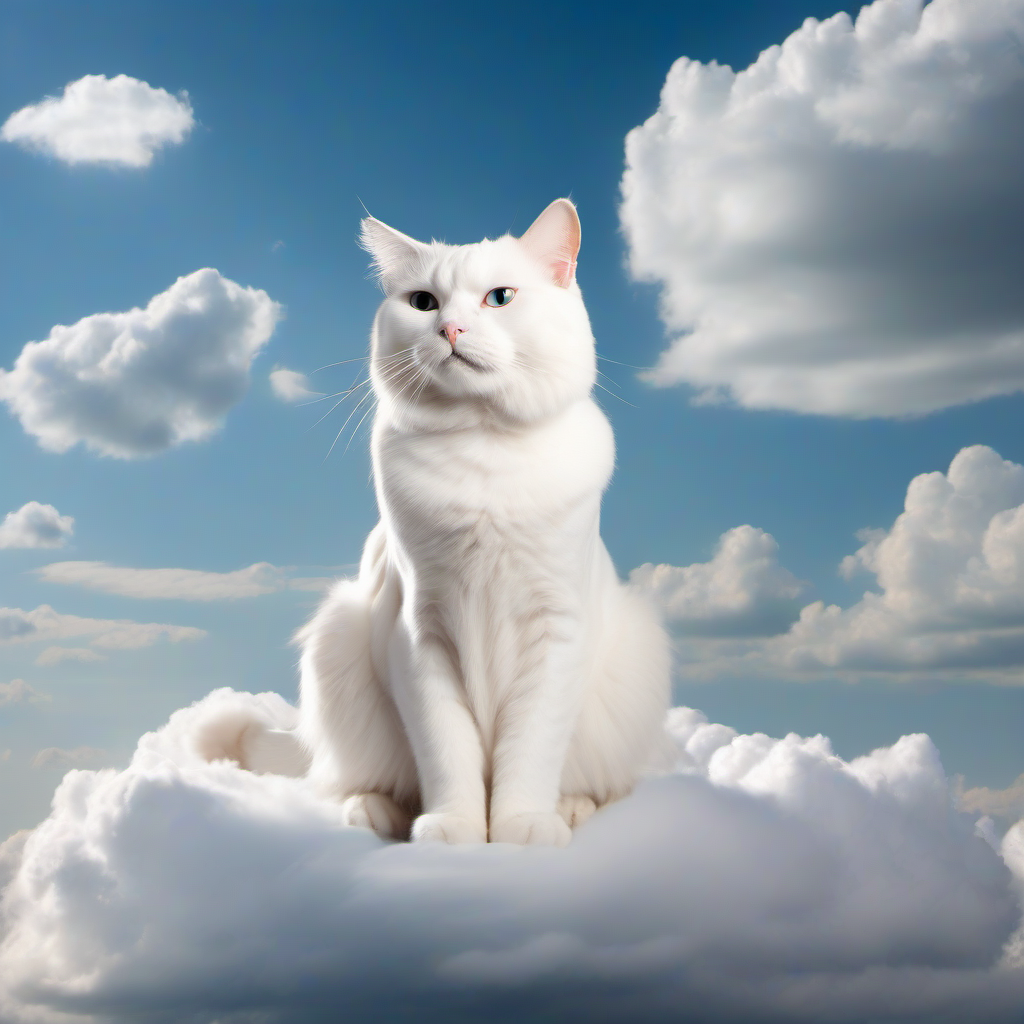}\caption{}\end{subfigure}
    \hfill
    \begin{subfigure}[b]{0.24\textwidth}\includegraphics[width=\linewidth]{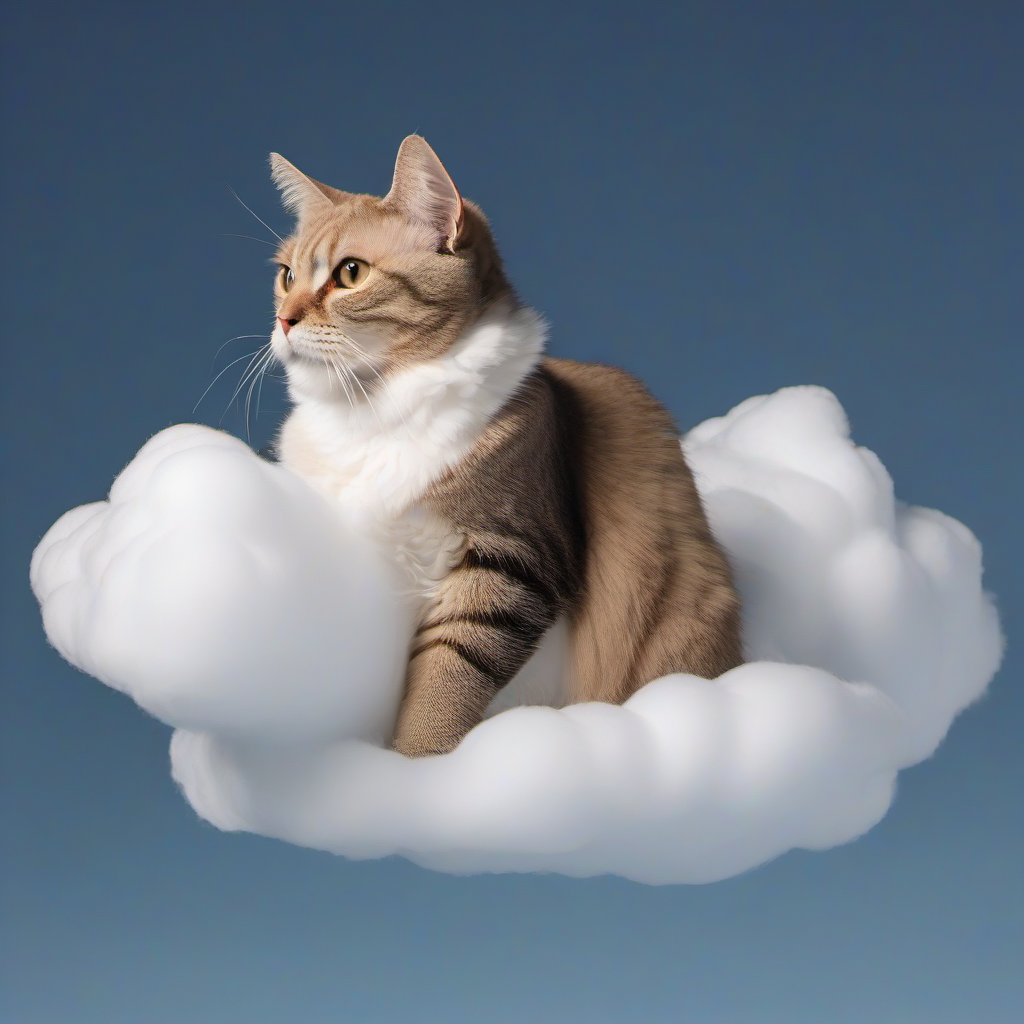}\caption{}\end{subfigure}
    \hfill
    \begin{subfigure}[b]{0.23\textwidth}\includegraphics[width=\linewidth]{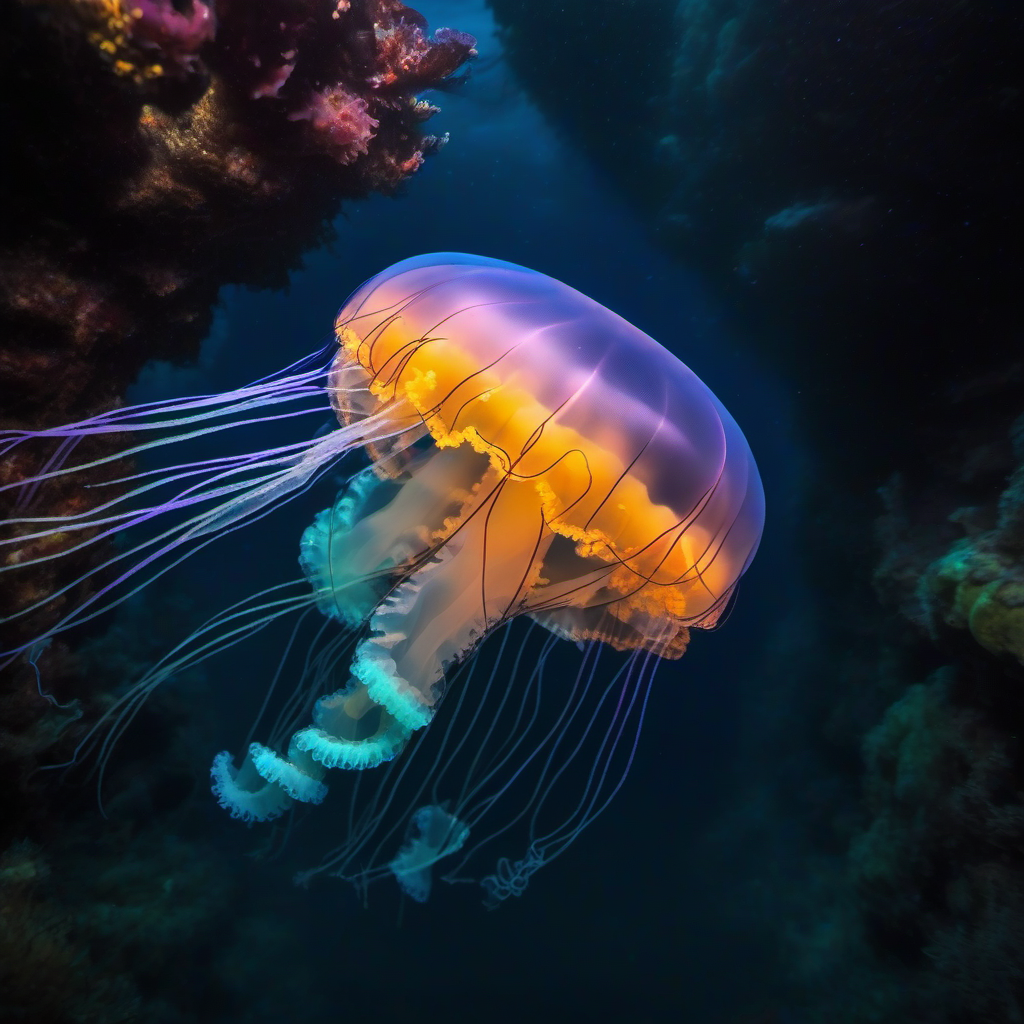}\caption{}\end{subfigure}
    }
    \resizebox{\linewidth}{!}{%
    \begin{subfigure}[b]{0.24\textwidth}\includegraphics[width=\linewidth]{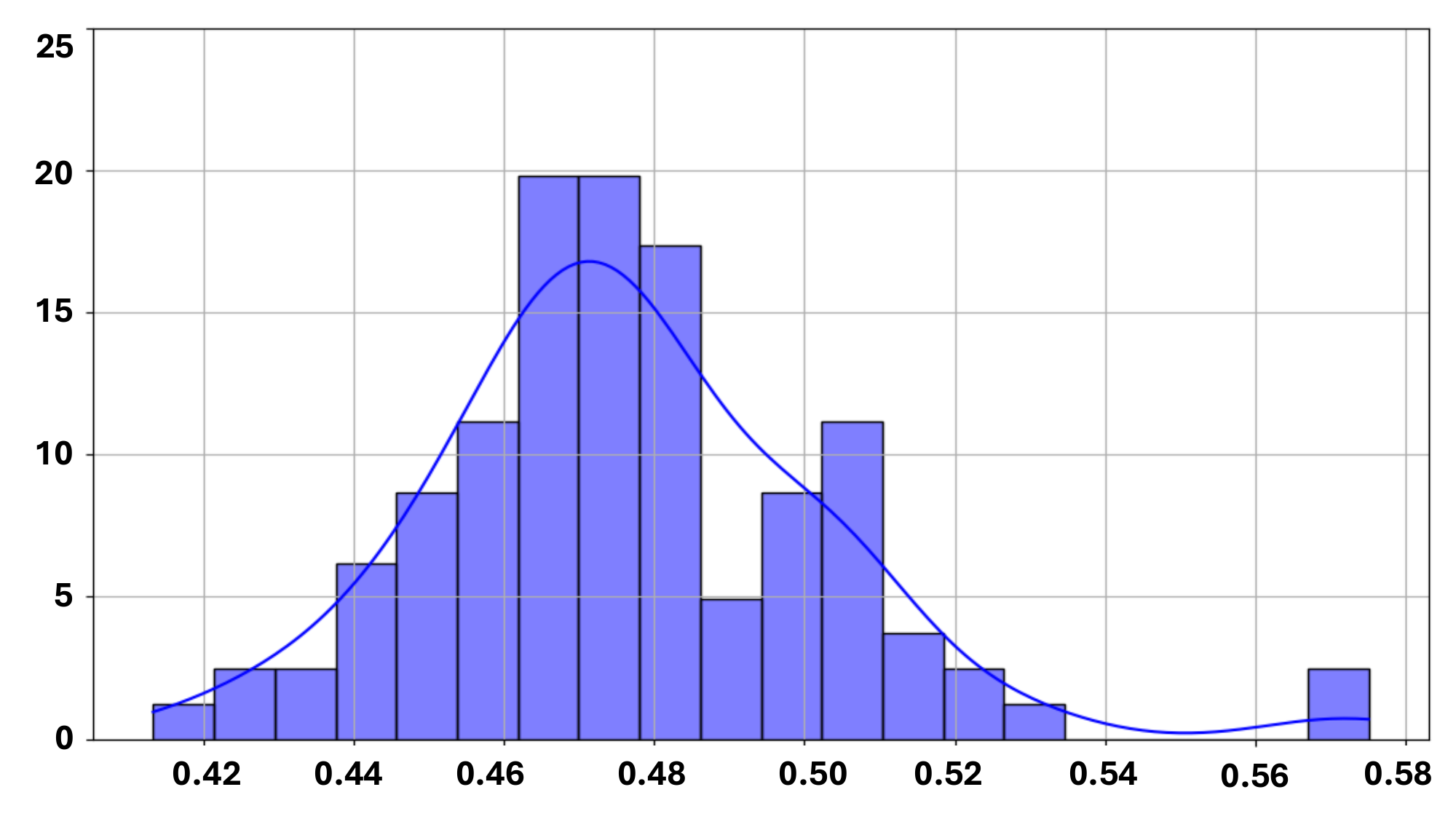}\caption{}\end{subfigure}
    \hfill
    \begin{subfigure}[b]{0.24\textwidth}\includegraphics[width=\linewidth]{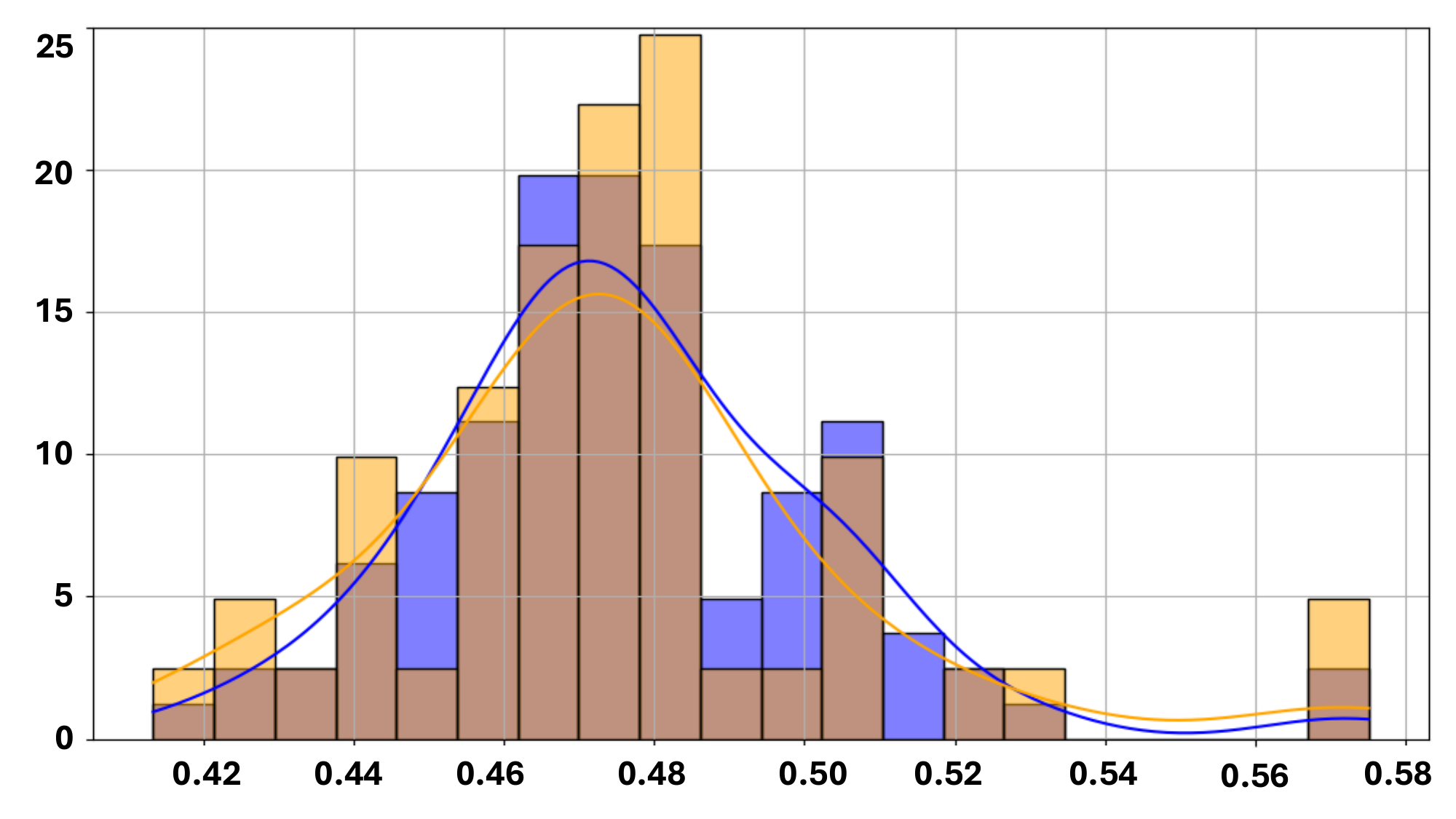}\caption{}\end{subfigure}
    \hfill
    \begin{subfigure}[b]{0.24\textwidth}\includegraphics[width=\linewidth]{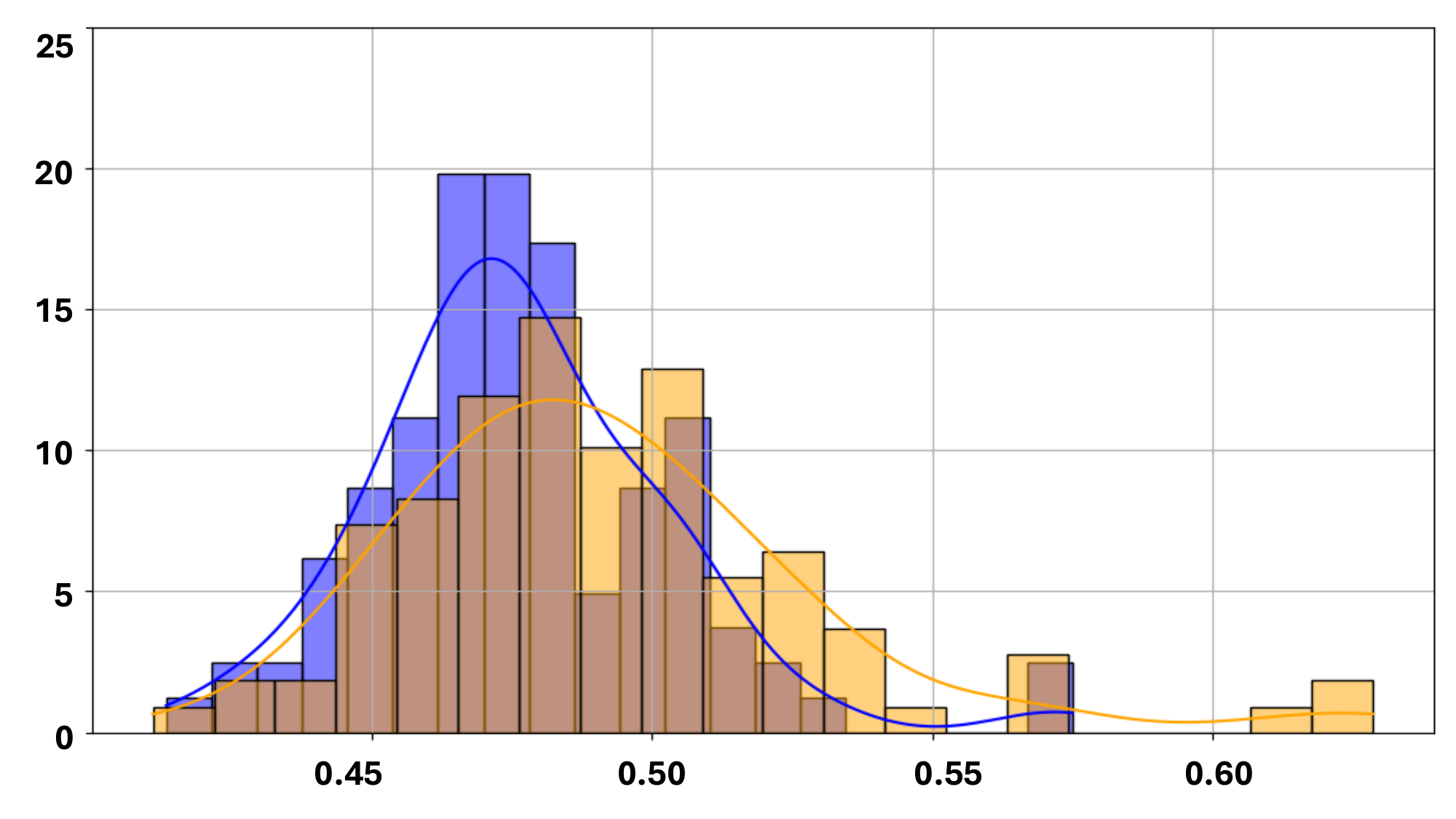}\caption{}\end{subfigure}
    \hfill
    \begin{subfigure}[b]{0.23\textwidth}\includegraphics[width=\linewidth]{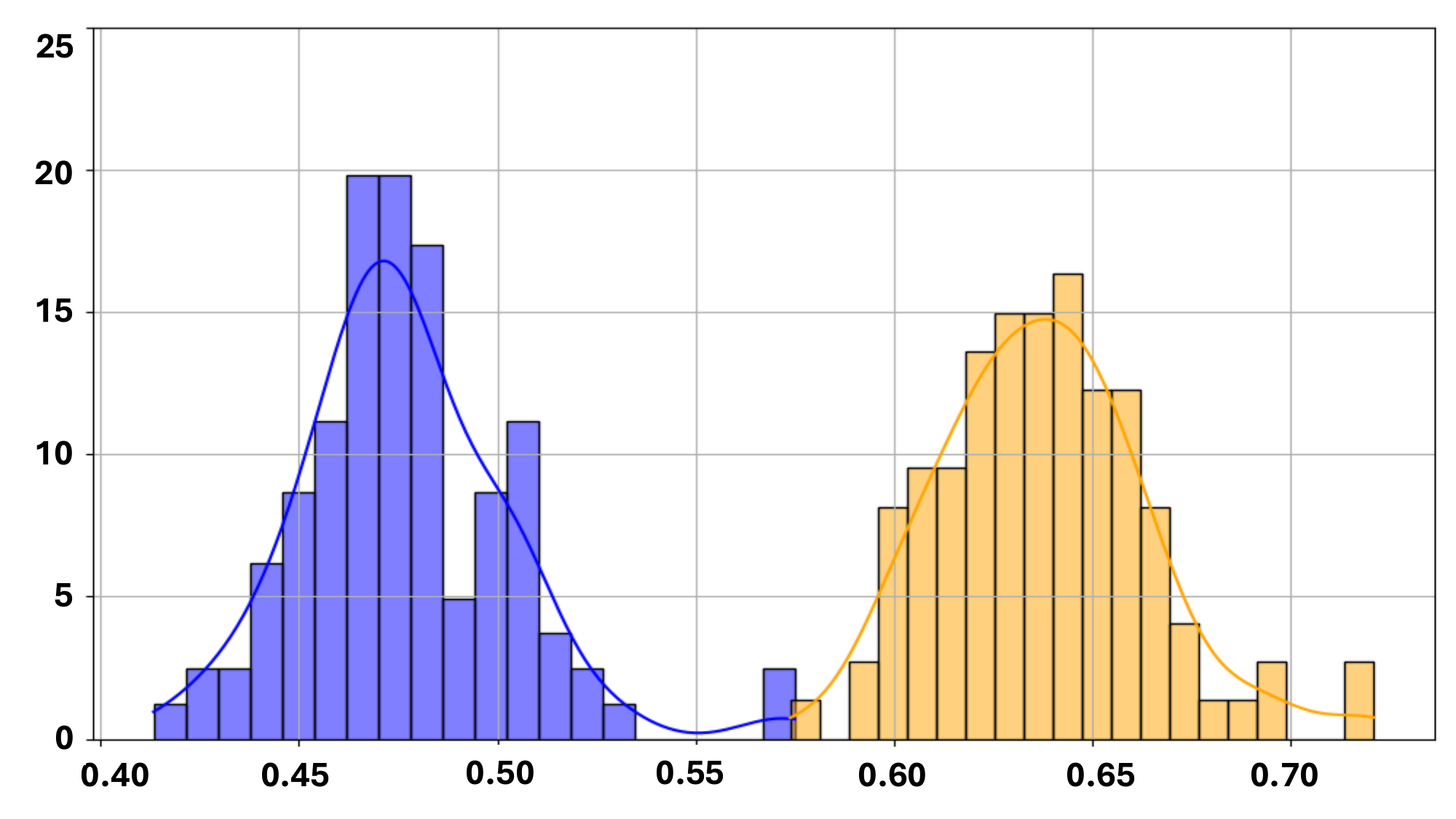}\caption{}\end{subfigure}
    }
    \caption{LPIPS distributions for images generated from different prompts using SDXL. (a,b) identical prompts; (c) less specific; (d) different subject. Divergence in (f--h) reflects decreasing similarity between inferred and reference prompts.}
    \label{fig:hit-miss-examples}
\end{figure}

\cref{fig:hit-miss-examples} illustrates LPIPS distributions for generations from varying prompt types. As prompt specificity decreases, similarity distributions diverge, indicating reduced correspondence between inferred and original prompts.

This distribution-aware evaluation extends prior single-image heuristics, offering a reproducible, model-agnostic framework for quantifying prompt inference success. Midjourney outputs were excluded from these analyses due to seed-control limitations inherent to its closed-source interface.

\subsection{Human–AI Combined Inference}
\label{sec:human-ai-expsetup}

To assess collaboration effects, we combined human-inferred prompts with AI-generated prompts from the CLIP Interrogator~\cite{clipinterrogator}, which produces descriptive text aligned with an input image’s CLIP embeddings. Each human–AI prompt pair was merged using GPT-4~\cite{openai2023gpt}, instructed to synthesize a single prompt of up to 25 words without adding new concepts. For example:

\begin{leftbar}
\small
\noindent
\textbf{Instruction:} Combine these two prompts into a 25-word description without extra details.\\
1. man dressed in steampunk in a steampunk factory\\
2. a man in a steampunk suit and top hat standing in front of a giant clock with gears, Bastien L. Deharme, fantasy art\\[1ex]
\textbf{GPT-4 Response:} a man in a steampunk suit and top hat stands in a factory surrounded by gears and clocks, embodying a fantasy art portrait.
\end{leftbar}

Capping prompt length ensured conciseness and prevented duplication of keywords that could bias generations. Using an LLM for merging, rather than simple concatenation, preserved linguistic coherence and balance between human and AI contributions. These combined prompts were later compared with human-only and AI-only prompts to evaluate synergy effects (see \cref{sec:colab-eval}).

\section{Results and Analysis}
\label{sec:results}

We report findings following the setup in \cref{sec:exp-setup}. We first measure human performance on prompt inference, then evaluate human-AI combined prompting, and finally analyze factors that influence success, including prompt type, model, and linguistic structure.

\subsection{RQ1: Human Performance on Prompt Inference Tasks}
\label{sec:human-eval}

For both the controlled and uncontrolled dataset, we quantify performance using the hit-rate criterion in \cref{sec:success-scores-expsetup} across ImageHash, LPIPS, and CLIP variants. Similarity score distributions from participant-inferred prompts are compared to those generated from the corresponding original prompt. These results are summarized in \cref{tab:hit-rates}.

\vspace{-4pt}
\begin{table}[!ht]
    \centering
    \begin{tabular}{l|c|c}
        \toprule
        \textbf{Similarity Metric} & \textbf{Controlled Hit Rate (\%)}  & \textbf{Uncontrolled Hit Rate (\%)}\\
        \midrule
        ImageHash     & 53.29   & 53.72 \\
        LPIPS         & 22.89   & 9.77 \\
        CLIP B32      & 7.28    & 7.56 \\
        CLIP L14      & 7.05    & 7.21 \\
        \bottomrule
    \end{tabular}
    \caption{Successful prompt inference (hits) by metric and target prompt set.}
    \label{tab:hit-rates}
\end{table}
\vspace{-20pt}

Hit rates differ by metric. ImageHash is highest (about 53\%) but captures shallow visual cues only and yields discrete distance values, which can inflate success and produce false positives relative to human judgment~\cite{trinh2025picture}. LPIPS is lower, especially for uncontrolled prompts, reflecting the difficulty of replicating fine perceptual details from open-ended images. CLIP metrics are lowest, indicating that matching semantic intent is hardest. Given these limitations, we exclude ImageHash from subsequent analyses and focus on LPIPS and CLIP.

\begin{leftbar}
\small
\textbf{H1} is partially supported. Participants often recover subjects but struggle with modifiers, which reduces LPIPS and CLIP hit rates. Differences between controlled and uncontrolled sets indicate that open-ended prompts increase difficulty.
\end{leftbar}

\subsection{RQ2: Impact of Prompt Linguistic Features and Model Choice}
\label{sec:prompt-model-eval}

We examine correlations between linguistic features, similarity statistics, hit outcomes, and model identity for CLIP and LPIPS.

\vspace{-20pt}
\begin{figure}[!ht]
    \centering
    \begin{subfigure}[b]{0.9\linewidth}
        \centering
        \includegraphics[width=\linewidth]{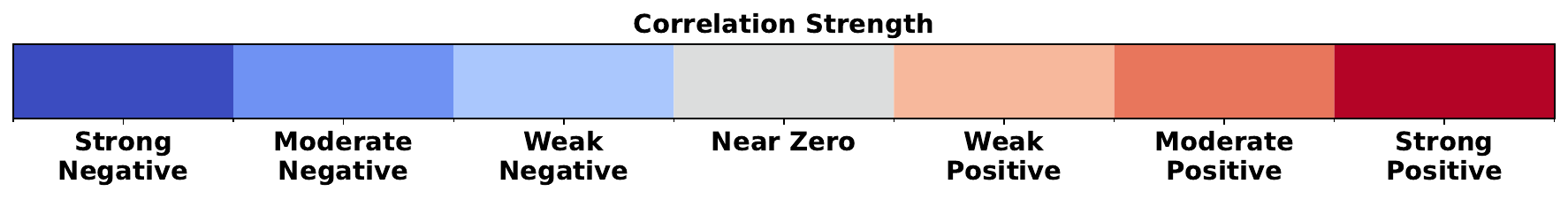}
    \end{subfigure}
    \begin{subfigure}[b]{0.49\linewidth}
        \centering
        \includegraphics[width=\linewidth]{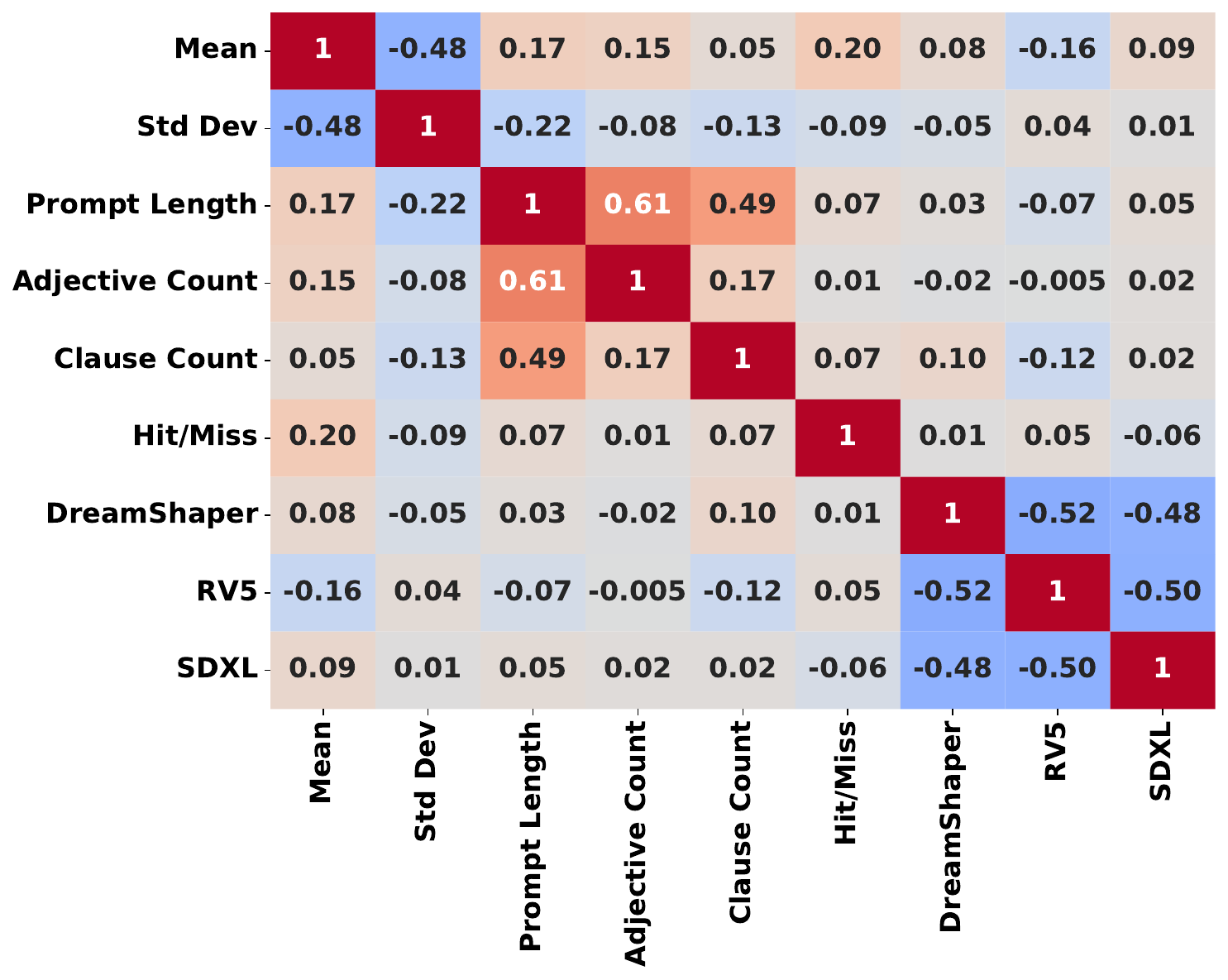}
        \caption{CLIP B32, controlled}
        \label{fig:clipb32-corr-allmodels-controlled}
    \end{subfigure}
    \hfill
    \begin{subfigure}[b]{0.49\linewidth}
        \centering
        \includegraphics[width=\linewidth]{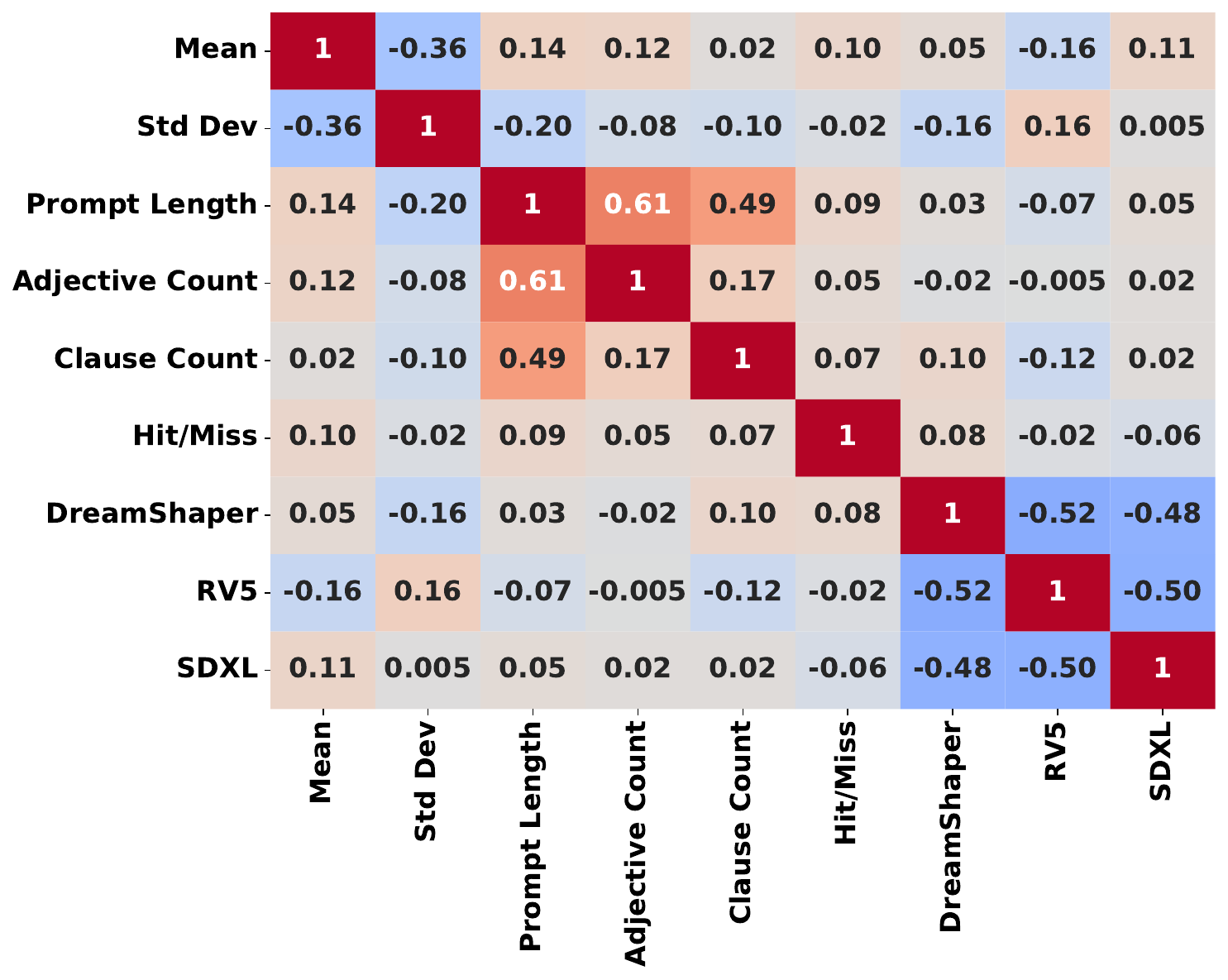}
        \caption{CLIP L14, controlled}
        \label{fig:clipl14-corr-allmodels-controlled}
    \end{subfigure}
    \begin{subfigure}[b]{0.49\linewidth}
        \centering
        \includegraphics[width=\linewidth]{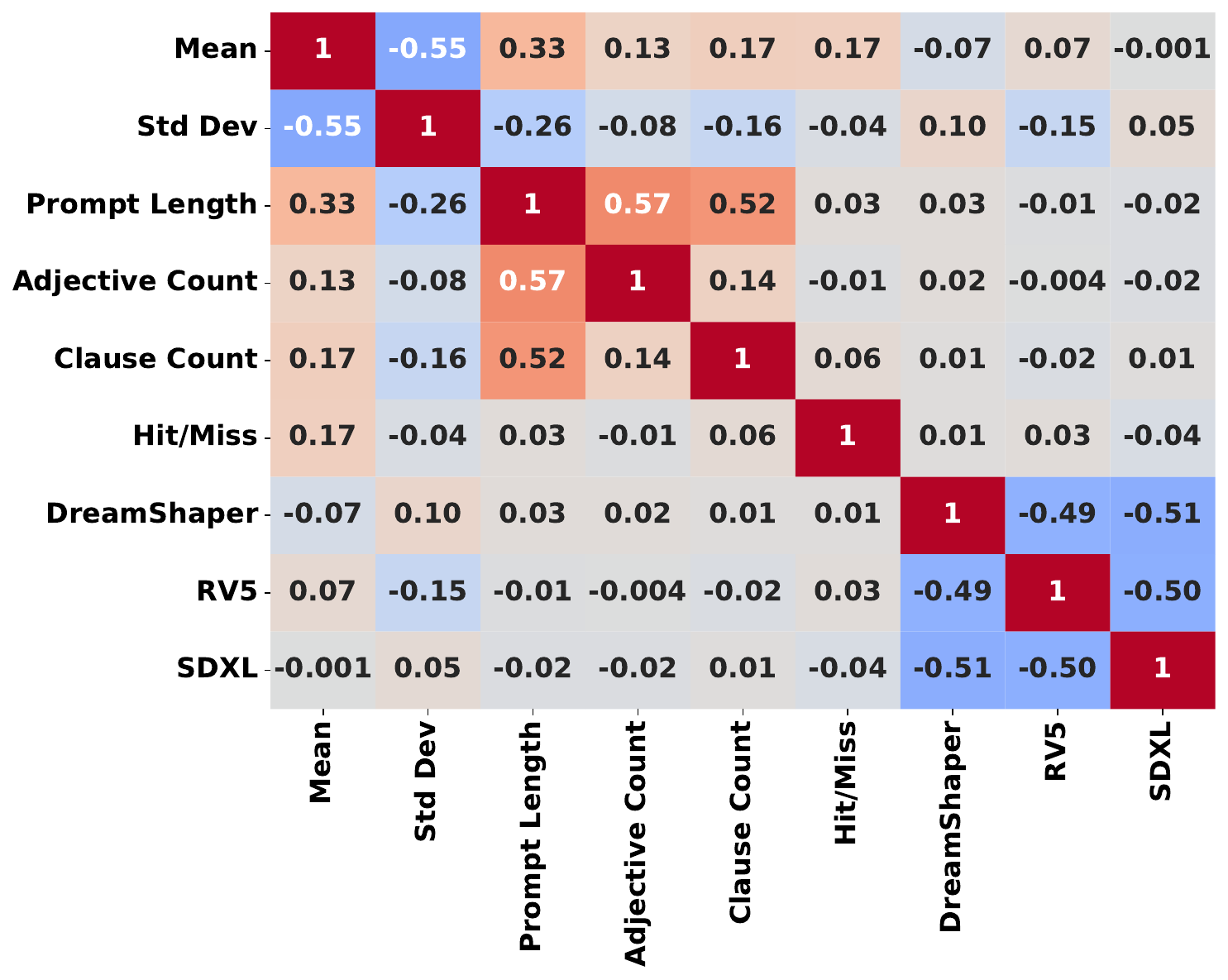}
        \caption{CLIP B32, uncontrolled}
        \label{fig:clipb32-corr-allmodels-uncontrolled}
    \end{subfigure}
    \hfill
    \begin{subfigure}[b]{0.49\linewidth}
        \centering
        \includegraphics[width=\linewidth]{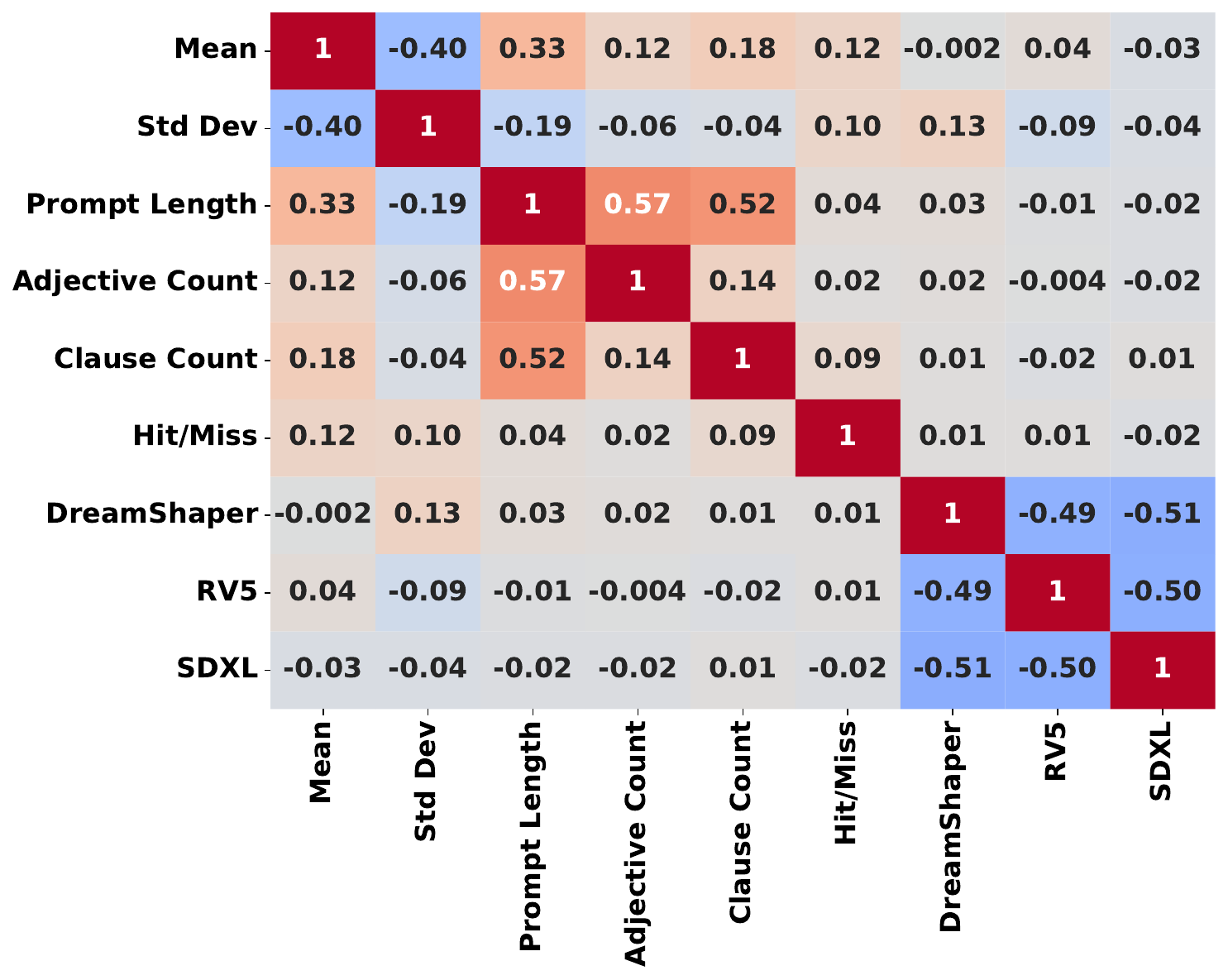}
        \caption{CLIP L14, uncontrolled}
        \label{fig:clipl14-corr-allmodels-uncontrolled}
    \end{subfigure}
    \caption{CLIP correlations with prompt features, similarity statistics, hit outcomes, and model identity. Binary variables use point-biserial associations. One-hot indicators are mutually exclusive.}
    \label{fig:clip-comparison}
\end{figure}

\cref{fig:clip-comparison} shows consistent linguistic patterns across CLIP variants: longer prompts correlate with more adjectives and clauses; mean CLIP has weak positive associations with success, while higher mean alignment tends to coincide with lower variance. Model effects are modest but consistent. DreamShaperXL associates with slightly better success and lower variability, while RV5 shows greater variability and weaker success. SDXL is largely neutral.

\begin{figure}[htbp]
    \centering
    \begin{subfigure}[b]{0.9\linewidth}
        \centering
        \includegraphics[width=\linewidth]{acm-t.web-figures/uncontrolled-prompts/comparison-across-models/heatmap_colorbar_banded_horizontal.pdf}
    \end{subfigure}
    \begin{subfigure}[b]{0.49\linewidth}
        \centering
        \includegraphics[width=\linewidth]{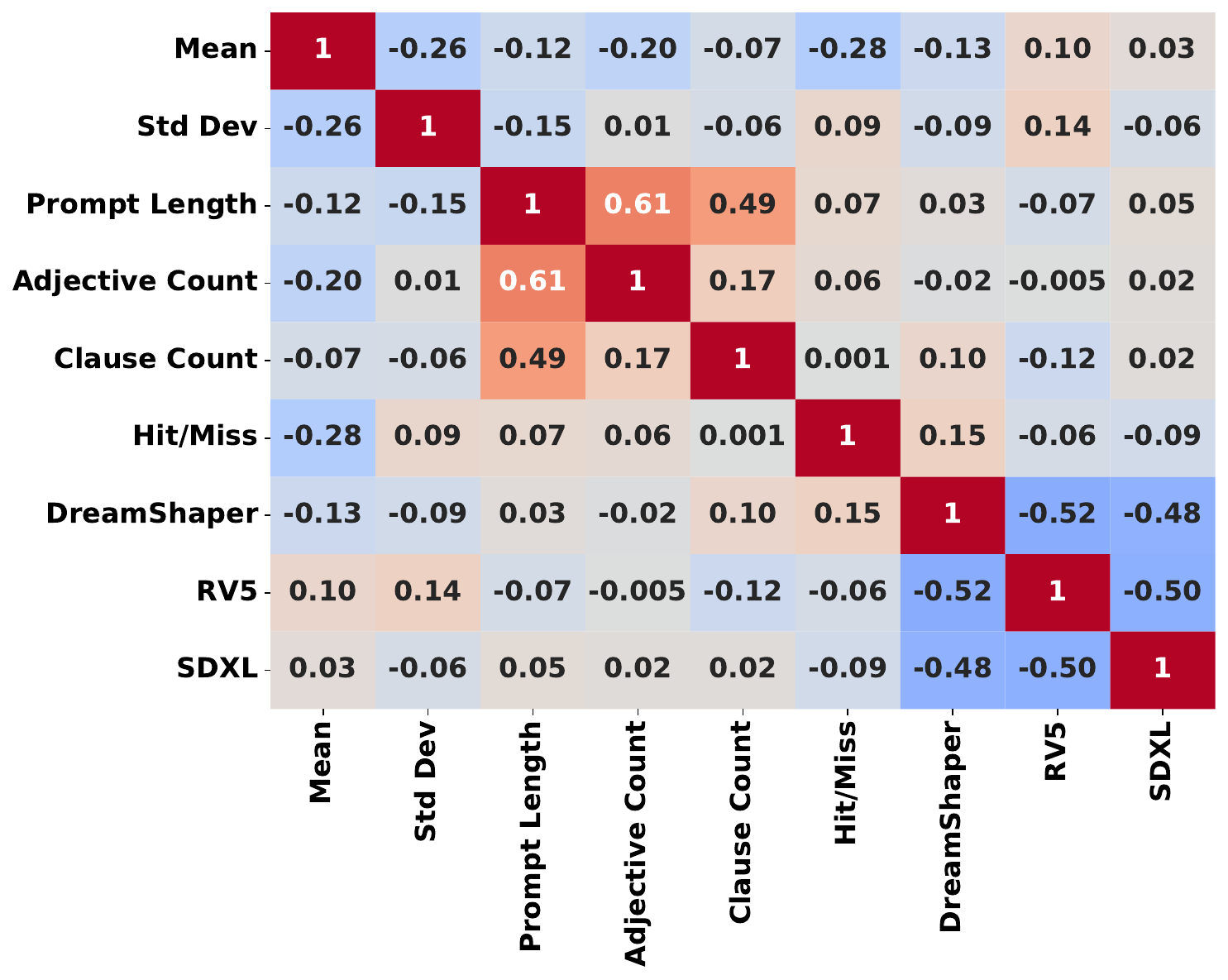}
        \caption{LPIPS, controlled}
        \label{fig:lpips-corr-allmodels-controlled}
    \end{subfigure}
    \hfill
    \begin{subfigure}[b]{0.49\linewidth}
        \centering
        \includegraphics[width=\linewidth]{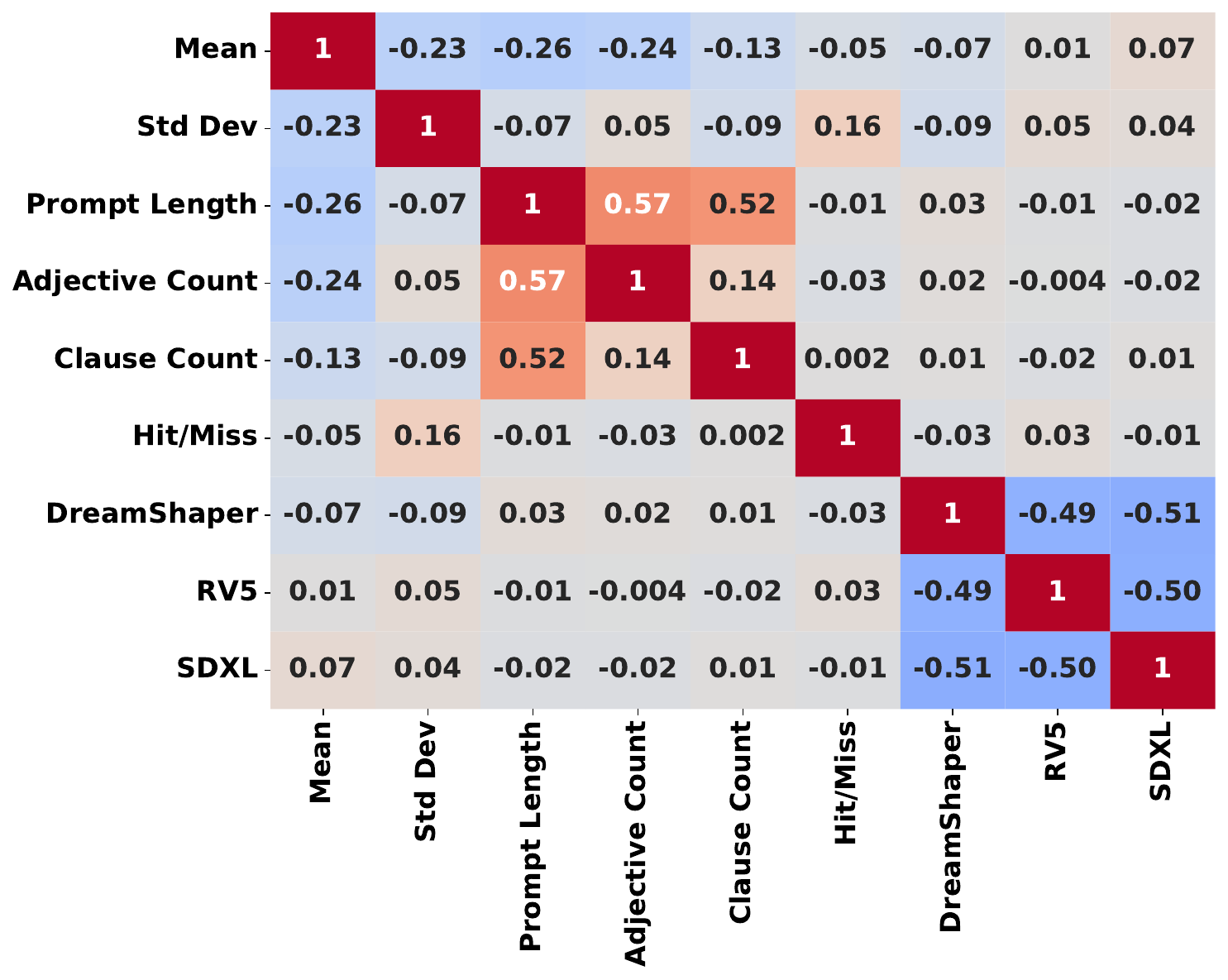}
        \caption{LPIPS, uncontrolled}
        \label{fig:lpips-corr-allmodels-uncontrolled}
    \end{subfigure}
    \caption{LPIPS correlations with prompt features, similarity statistics, hit outcomes, and model identity.}
    \label{fig:lpips-comparison}
\end{figure}

For LPIPS (\cref{fig:lpips-comparison}), mean LPIPS correlates negatively with success (controlled $r=-0.28$, uncontrolled $r=-0.05$), so lower LPIPS relates to more hits. Mean LPIPS also has small negative correlations with length, adjectives, and clauses. LPIPS variance has weak positive correlation with success (controlled $r=0.09$, uncontrolled $r=0.16$). DreamShaperXL shows the strongest positive relation with success (controlled $r=0.15$) and the most negative relation with mean LPIPS (controlled $r=-0.13$, uncontrolled $r=-0.07$). RV5 tends toward higher variability (controlled $r=0.14$, uncontrolled $r=0.05$).

\begin{leftbar}
\small
\textbf{H2} is partially supported. Linguistic richness and model choice have modest effects. DreamShaperXL generations are more stable and perceptually aligned, which aids inference. RV5 is more variable. Trends are consistent across controlled and uncontrolled sets.
\end{leftbar}

\subsection{RQ3: Human-AI Combined Prompt Inference}
\label{sec:colab-eval}

We compare hit rates for human-only prompts with human-AI combined prompts formed via CLIP Interrogator plus GPT-4 merging (\cref{sec:human-ai-expsetup}). Results are in \cref{tab:human-ai-hit-rates}. ImageHash is reported for completeness. Combined prompts improve ImageHash but not LPIPS or CLIP. This indicates better low-level resemblance without improved perceptual or semantic alignment.

\vspace{-8pt}
\begin{table}[htbp]
\centering
\resizebox{\linewidth}{!}{%
\begin{tabular}{l|c|c}
\toprule
\textbf{Similarity Metric} &
\textbf{\shortstack{Human Only Hit Rate (\%)}} &
\textbf{\shortstack{Human--AI Combined Hit Rate (\%)}}\\
\midrule
ImageHash & 53.29 & 60.62 \\
LPIPS & 22.89 & 18.59 \\
CLIP B32 & 7.28 & 5.89 \\
CLIP L14 & 7.05 & 5.77 \\
\bottomrule
\end{tabular}
}
\caption{Hit rates for human-only and human--AI combined prompts.}
\label{tab:human-ai-hit-rates}
\end{table}

\begin{leftbar}
\small
\textbf{H3} is not supported. The current merging strategy does not outperform human-only prompting on perceptual or semantic metrics. Future collaborative methods should target alignment with LPIPS and CLIP, not shallow similarity.
\end{leftbar}

\subsection{Discussion of Key Findings}

Humans can often infer core subjects, especially in controlled settings, but struggle to recover modifiers and stylistic intent. Metric choice matters: ImageHash is optimistic and misaligned with human judgment~\cite{trinh2025picture}, while LPIPS and CLIP are stricter and reveal lower success for open-ended prompts. Linguistic richness and model choice show modest but consistent effects; DreamShaperXL tends to produce more stable, aligned generations that are somewhat easier to approximate. Human-AI prompt merging increased shallow similarity only and reduced LPIPS and CLIP performance. Overall, results suggest that precise semantic alignment remains difficult in unconstrained contexts, which provides some resilience for prompt-based intellectual property.

\section*{Conclusion}
\label{sec:conclusion}

This study presents a human-subject evaluation of prompt inference in AI-generated images, with implications for content authorship, web-based generative tools, and the emerging economy of prompt marketplaces. Our findings show that while human participants can infer broad subject matter from an image, they often struggle with accurately reconstructing modifiers and stylistic intent, particularly in unconstrained contexts. This limitation becomes more apparent when evaluated with perceptual and semantic similarity metrics such as LPIPS and CLIP, which better reflect alignment with the original prompt.

Notably, human-AI combined prompts did not outperform human inferences alone. Instead, na\"\i ve merging strategies often diluted semantic coherence, resulting in lower similarity scores under meaningful evaluation metrics. These findings suggest that while generative tools enable accessible creativity, effective collaboration between human users and AI systems remains a design challenge. Moreover, participant background, including AI familiarity, had limited impact on success rates, implying that prompt inference performance is more strongly influenced by task and system-level constraints than by user expertise.

Overall, these results provide evidence that prompts are not easily reverse-engineered from generated outputs. This supports the relative resilience of prompt-based intellectual property in current generative workflows. For web-based platforms hosting generative models and prompt-sharing ecosystems, this raises key considerations around content provenance, authorship claims, and secure prompt design.

\bibliographystyle{splncs04}
\bibliography{references}

\end{document}